\documentclass{aa}
\usepackage{txfonts}
\usepackage{graphicx}
\usepackage{natbib}

\def\solar {\ifmmode_{\mathord\odot} \else $_{\mathord\odot}$\fi}
\def\Msol {\ifmmode {\,{\it M}\solar} \else $\,M$\solar\fi}     
\def\Rsol {\ifmmode {\,{\it R}\solar} \else $\,R$\solar\fi}     
\def\Lsol {\ifmmode {\,{\it L}\solar} \else $\,L$\solar\fi}     
\newcommand{\Mjup}{M$_{Jup}$}
\newcommand{\Mearth}{M$_{Earth}$}
\newcommand{\Mnep}{M$_{Nep}$}

\newcommand{\periode}{5.366}     
\newcommand{\eperiode}{0.001}   
\newcommand{\T}{51004.30}          
\newcommand{\eT}{0.06}                 
\newcommand{\e}{0. (fixed)}        
\newcommand{\V}{-9.212}                
\newcommand{\eV}{0.001}    
\newcommand{\w}{0. (fixed)}            
\newcommand{\K}{13.2}                    
\newcommand{\eK}{0.4}   
\newcommand{\aunsini}{0.652\,10$^{-2}$} 
\newcommand{\mdeuxsini}{16.6}           
\newcommand{\mdeuxsinijup}{0.052}      
\newcommand{\mdeuxsininep}{0.97}     
\newcommand{\demiaxe}{0.041}                  
\newcommand{\sigmaOmC}{2.5}     

\begin{document}

\title{The HARPS search for southern extra-solar planets
       \thanks{Based on observations made with the HARPS instrument on the 
         ESO 3.6-m telescope at La Silla Observatory under programme ID 
         072.C-0488}
       } 

\subtitle{VI. A Neptune-mass planet around the nearby M dwarf Gl~581}

\author{X.~Bonfils \inst{1,2} 
   \and T.~Forveille \inst{3,1}
   \and X.~Delfosse \inst{1}
   \and S.~Udry \inst{2}
   \and M.~Mayor \inst{2}
   \and C.~Perrier \inst{1}\\
   F.~Bouchy \inst{4}
   \and F.~Pepe \inst{2}
   \and D.~Queloz \inst{2}
   \and J.-L.~Bertaux \inst{5} 
}
\institute{Laboratoire d'Astrophysique, Observatoire de Grenoble, BP 53, 
           F-38041 Grenoble, C\'edex 9, France
    \and
     Observatoire de Gen\`eve, 51 ch. des Maillettes, 1290 Sauverny, 
           Switzerland
     \and
     Canada-France-Hawaii Telescope Corporation, 65-1238 Mamalahoa Highway, 
           Kamuela, HI96743, Hawaii, USA
     \and
     Laboratoire d'Astrophysique de Marseille, Traverse du Siphon, 
           13013 Marseille, France
     \and
     Service d'A\'eronomie du CNRS, BP 3, 91371 Verrières-le-Buisson, France
   }

\date{Received / Accepted}

\abstract{We report the discovery of a Neptune-mass planet around Gl~581 
(M3V, M~=~0.31\Msol), based on precise Doppler measurements with the HARPS 
spectrograph at La Silla Observatory. The radial velocities reveal a 
circular orbit of period P~=~\periode~days and semi-amplitude 
K$_1$~=~\K~m~s$^{-1}$. The resulting minimum mass of the planet 
($m_2 \sin{i}$) is only 0.052~\Mjup~=~0.97~\Mnep~=~16.6~\Mearth~ 
making Gl~581b one of the lightest extra-solar planet known to date. 
The Gl~581 planetary system is only the third centered on an M~dwarf, 
joining the Gl~876 three-planet system and the lone planet around Gl~436. Its 
discovery reinforces the emerging tendency of such planets to be of low 
mass, and found at short orbital periods. The statistical properties of
the planets orbiting M~dwarfs do not seem to match a simple mass scaling
of their counterparts around solar-type stars.

}

\titlerunning{A Neptune-mass planet around the nearby M dwarf Gl~581}
\authorrunning{Bonfils et al.}

\maketitle

\section{Introduction}
Over 150 planets have been found orbiting main sequence stars other 
than the Sun, in about 140 planetary systems of which 18 have multiple
planets (http://vo.obspm.fr/exoplanetes/encyclo/).
These extra-solar planets are a very diverse class: their mass ranges
between half the mass of Neptune and 15 times the mass of Jupiter, some 
have large eccentricities when others have nearly circular orbits, 
their periods range from slightly over a day 
\citep[OGLE-TR-56]{2003Natur.421..507K} to
over a decade \citep[55~Cnc]{2002ApJ...581.1375M}. The multiple 
systems range from strongly resonant to fully hierarchical 
\citep{Rivera2005,2002ApJ...581.1375M}. This diversity demonstrates
that our own solar system represents but one possible outcome of the
planetary formation and evolution processes, and apparently not even
a very common one.

The statistical properties of these exoplanets provide crucial clues
to their formation mechanism. As perhaps the most dramatic example,
the seminal detection of the 51~Peg planet in a 4-days orbit 
\citep{1995Natur.378..355M} immediately forced theoreticians to 
recognize the critical importance of orbital migration 
\citep{1996Natur.380..606L,1997Icar..126..261W}.
The correlation between the occurence of Jupiter-mass planet and the 
high metallicity of the host stars \citep{gonzalez97,santos01,santos04}
is another exemple. It is thought to reflect the controlling role 
of the condensate mass in the protoplanetary disk, but it has taken 
longer to converge towards that consensus.

To date, all but 2 of these 140 planetary systems orbit solar-type stars. 
In part, this no doubts reflects a bias of most planet-search programmes 
towards the relatively bright F to K main sequence stars, and away from their 
fainter M-type counterparts (M~$<$~0.6\Msol). Nonetheless, several teams
\citep{2004ApJS..152..261W,2003AJ....126.3099E,1998A&A...338L..67D} 
collectively monitor over 200~M dwarfs with
sufficient precision to detect a Jupiter-mass planet out to at least 2~AU.
These efforts have up to now identified the 3-planet system around Gl~876 
\citep{1998A&A...338L..67D,1998ApJ...505L.147M,2001ApJ...556..296M,Rivera2005},
and the single-planet Gl~436 system  \citep{2004ApJ...617..580B}. Of
these 4 planets, 2 are in the Neptune-mass class, leaving only two of 
the Gl~876 planets with approximately Jupiter-mass. By constrast
$\geq$5\% of solar-type stars have Jupiter-mass planets 
\citep{2000prpl.conf.1285M}, and the comparative deficit for the
M dwarfs is therefore statistically robust \citep{2004ApJ...617..580B,naef05}.

An open question, though, is whether M dwarfs genuinely have fewer planets,
or whether their planets are just as abundant, but not quite as massive.
Addressing this question needs higher precision, or more measurements,
than the radial-velocity surveys have achieved to date. To help answering
this question, we are using the HARPS spectrograph for a high-precision survey 
of more than 100 nearby M dwarfs. We present here its first detection, a Neptune-mass planet around Gl~581.

\section{Properties of Gl~581}
Gl~581 (HIP~74995,  LHS~394) is an M3 dwarf  \citep{1997AJ....113.1458H}
with a distance to the Sun of 6.3~pc ($\pi$~=~159.52$\pm$2.27, 
\cite{1997yCat.1239....0E}).
Its photometry (V=10.55$\pm$0.01, B$-$V=1.60 \citep{1997A&AS..124..349M}; 
K=5.85$\pm$0.03 \citep{1992ApJS...82..351L}) and the parallax together result
in absolute magnitudes of $M_V$=11.56$\pm$0.03
and $M_K$=6.86$\pm$0.04. 
From its absolute V magnitude and the 2.08 V-band bolometric correction of 
\cite{1998A&A...331..581D}, the luminosity of Gl~581 is 0.013~\Lsol.
The \cite{2000A&A...364..217D} K-band mass-luminosity relation, which
has much lower intrinsic dispersion than the equivalent V-band relation,
gives a 0.31$\pm$0.02\Msol mass, 
and the \cite{2000ARA&A..38..337C} theoretical Mass-Radius relation 
then a radius of 0.29~\Rsol. Interestingly, \cite{Bonfils2005} find Gl~581
slightly metal-poor ([Fe/H]=$-$0.25), in contrast to most planet-host stars 
having supersolar metallicities.

Gl~581 has been classified as a variable star (HO Lib). However, the data which have led to this classification \citep{weis94} have a short-term variability of $\sim$ 0.006 mag. The variability quoted by the author is marginally above the errors bars and, if real, has most likely a long-term nature
(several years).

\begin{figure}
\centering
\includegraphics[width=9cm]{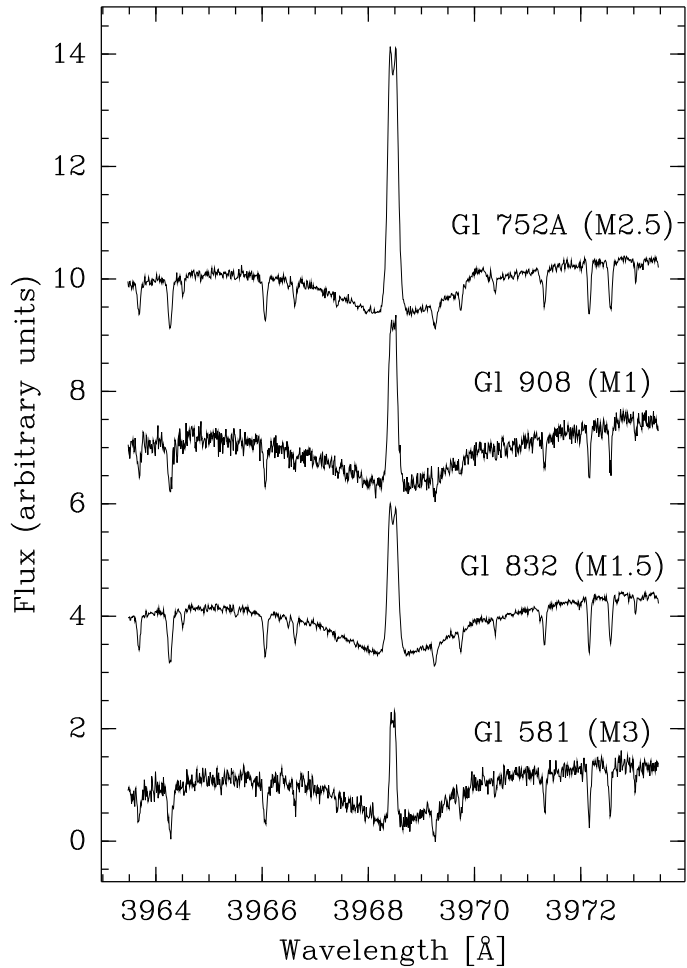}
	\caption{HARPS spectra of the \ion{Ca}{II} H 
($\lambda$~=~3968.47~$\AA$) line region for Gl\,581 and three comparison 
stars with similar spectral type and apparent magnitude. The stars are
displayed in order of ascending chromospheric activity, and from top to 
bottom are Gl\,752A, Gl\,908, Gl\,832 and  Gl\,581. The chromospheric
emission peaks look prominent against the weak blue continuum of these
M dwarfs, but they actually denote very weak chromospheric emission relative
to the bolometric luminosity. Amongst those 4 stars, Gl\,581 has the 
weakest chromospheric activity.
}
	\label{caII}
\end{figure}

The age of Gl~581 can be estimated from its kinematic characteristics, its
magnetic activity, and its metallicity, all of which point towards a 
moderate to older age. \cite{1992ApJS...82..351L} find that its UVW galactic
velocities are intermediate between those typical of the young and old
galactic disk, and \cite{1998A&A...331..581D} find very low X-ray emission
($L_x/L_{bol}<5.10^{-6}$) and a 2.1~km~s$^{-1}$ upper limit on the projected
rotation velocity, v~sini. 
The HARPS spectra show weak  Ca$_{II}$ H and K emission, in
the lower quartile of stars with similar spectral types (Fig.~\ref{caII}).
As mentioned above, Gl~581 also has a subsolar metallicity. 
Altogether, these properties suggest that it is at least 2~Gyr old, and 
they ensure that the radial velocity ``jitter'' from magnetic activity 
must be minimal.

\section{Doppler measurements and orbital analysis}
HARPS (High Accuracy Radial velocity Planet Searcher) is the new ESO 
high-resolution (R~=~115 000) fiber-fed echelle spectrograph, optimised
for planet search programmes and asteroseismology. It has proved to be 
the most precise spectro-velocimeter to date, reaching an instrumental 
RV accuracy better than 1~m~s$^{-1}$ 
\citep{2003Msngr.114...20M,2004A&A...426L..19S,2004A&A...423..385P,lovis05},
and even better on the short-term scales of interest for 
asteroseismology. For ultimate radial velocity precision HARPS uses 
simultaneous exposures of a thorium lamp through a calibration fiber.
When observing M dwarfs however, we rely instead on its very high 
instrumental stability (nightly instrumental drifts $<$~1~m~s$^{-1}$).
The M dwarfs are typically too faint for us to reach the stability limit
of HARPS within realistic integration times, and dispensing with the
simultaneous thorium light produces much cleaner stellar spectra, suitable
for quantitative spectroscopic analyses.

For the V~=~10.5 Gl~581 we use 15~mn exposures, and the median S/N ratio 
of our 20 spectra is 40 per pixel at 550~nm. The radial 
velocities (Table~\ref{TableRV}, only available electronically) were
obtained with the standard HARPS reduction pipeline, based on the 
cross-correlation with a stellar template and the precise nightly 
wavelength calibration with ThAr spectra \citep{1996A&AS..119..373B}.
They have a median internal error of only 1.3~m/s, which includes both 
the nightly zero-point calibration  uncertainty ($\sim$~0.8~m s$^{-1}$)
and the photon noise, computed from the full Doppler
information content of the spectra \citep{2001A&A...374..733B}.

The computed velocities exhibit an rms dispersion of 10~m~s$^{-1}$, much above 
their internal errors and also considerably more than we observe for 
stars with higher chromospheric activity. Of the three comparison stars 
with stronger chromospheric emission in Fig.~\ref{caII}, Gl\,752 and 
Gl\,832 have enough HARPS radial velocities to measure rms dispersions 
of 1.2 m/s (from 10 measurements) and 1.8 m/s (from 19 measurements). 
\cite{Rivera2005} report an rms dispersion of 3.42 m/s for the the third,
Gl\,908, dominated by their measurement noise. 

\begin{figure}
\centering
\includegraphics[width=9cm]{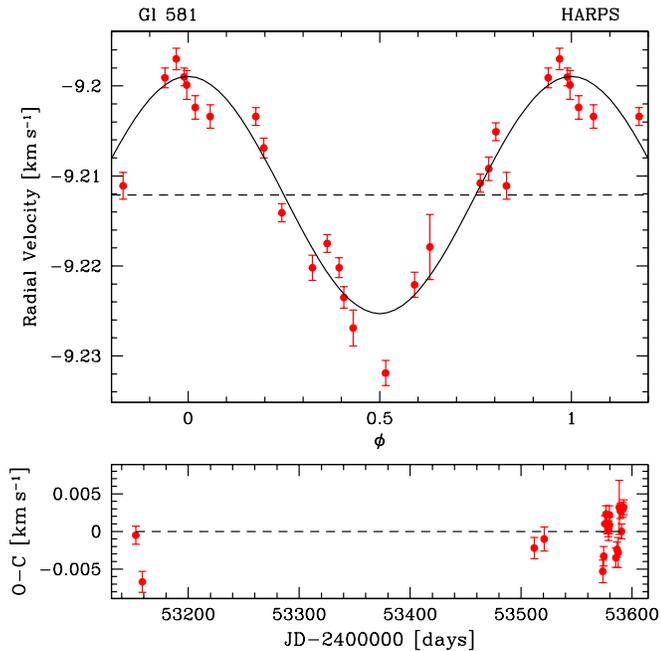}
	\caption{{\it Upper panel:} Phased radial velocities for Gl\,581. {\it Lower panel:} Residuals around the fitted solution versus time.}
	\label{phaseomc}
\end{figure}

\begin{figure}
\centering
\includegraphics[width=8cm]{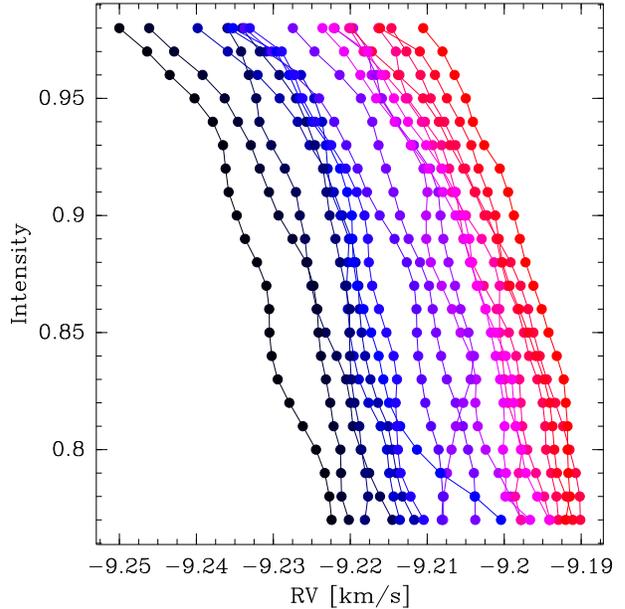}
	\caption{Bissectors of the HARPS correlation profiles for Gl~581.
         The shape of the bissector curve does not change together with its
         position. It validates the planetary interpretation as the cause of the radial-velocity variation.
}
	\label{bissector}
\end{figure}

As demonstrated by Fig.\ref{phaseomc}, a circular orbit of period P~=~{\periode}~d
and semi-amplitude K~=~{\K} m~s$^{-1}$ is an excellent fit to these
velocities, as expected from rapid tidal circularisation 
at this short period. Attempts at adjusting elliptical orbits resulted in 
non-significant eccentricities. We therefore adopt a circular orbit. Its parameters 
are listed in Table~\ref{orbparam} and lead to a minimum mass ($M~sin~i$) 
for the planet 
of only \mdeuxsinijup~\Mjup~=~\mdeuxsininep~\Mnep~=~\mdeuxsini~\Mearth. 
The weighted rms of the residuals around the fit is 2.5~m~s$^{-1}$, and
twice the internal errors of the measurements. More data points are needed to 
establish whether this extra dispersion is intrinsic to the star, or 
whether it could be explained by the presence of a third body in the 
system. 
Further credit is given to the latter hypothesis by the very-low variability level in the shape of the bisector (Fig~\ref{bissector}).
An overall translation dominates the evolution of the spectral profile,
leaving no doubt on the keplerian origin of the radial-velocity variations.
We also have available 11 ELODIE spectra, which extend the measurement 
baseline to 9~years, albeit with a 5~years gap between mid-2000 and 
mid-2005. Their 17~m/s median error bars are too 
large to reveal the K=13~m~s$^{-1}$ planetary signal.
They exhibit an rms dispersion of 23 m~s$^{-1}$ and lack 
any obvious long term trend, demonstrating that the Gl~581 system 
contains no Jupiter-mass planet with any period shorter than $\approx$10~years. 

\begin{table}
\caption{Orbital and physical parameters.}
\label{orbparam}
\centering
\begin{tabular}{l l c }
\hline\hline
\multicolumn{2}{l}{\bf Parameter} & \bf Gl\,581 b \\
\hline
$P$                 & [days]                 & \periode~$\pm$~\eperiode \\
$T$                 & [JD-2400000]   & \T~$\pm$~\eT \\
$e$                 &                             & \e \\
$V$                 & [km s$^{-1}$]    & \V~$\pm$~\eV \\
$\omega$      & [deg]                  & \w \\
$K$                 & [m s$^{-1}$]      & \K~$\pm$~\eK \\
$a_1 \sin{i}$  & [10$^{-3}$ AU] & \aunsini \\
$f(m)$             & [10$^{-9} M_{\odot}$] & 0.1354\,10$^{-2}$ \\
$m_2 \sin{i}$ & [$M_{\mathrm{Earth}}$] & \mdeuxsini \\
$a$ & [AU]     & \demiaxe \\
\hline
$N_{\mathrm{meas}}$ & & 20 \\
$Span$ & [days] & 440\\
$\sigma$ (O-C) & [m s$^{-1}$] & \sigmaOmC \\
\hline
\end{tabular}
\end{table}

\section{Discussion}
The semi-major axis of the planetary orbit is only 0.042~AU or 9 solar radii, similar
to most close-in planets around solar-type stars. In the natural length unit 
of its central star however, this amounts to 31 radii of Gl~581. The 
geometric transit probability is thus only 3\%, and significantly less than 
the $\approx$10\% typical of close-in planets around solar-type stars. If 
transits do occur on the other hand, the planet will cover a larger fraction 
of its smaller star. For a constant planetary radius, transits would thus 
be correspondingly deeper and more easily detected. At this
radius and given the 0.013~\Lsol~luminosity of the star, the expected 
temperature of the planetary surface is $\approx$420~K, with large 
uncertainties from the unknown albedo and energy transport. Even with
conservative error bars though, this temperature is compatible with either
a rocky planet or a gas giant, and evaporation will be negligibly small
in either configuration.

The detection of Gl~581b brings the inventory of M-dwarf which harbor 
planetary systems to 3, and the number of their planets to 5. While 
admittedly still very small, these samples allow us an initial peek at their 
properties as a population, compared to the much more numerous planets 
known around solar-type stars. One immediate observation is that none of 
the three stars is metal-rich, with Gl~876, Gl~436 and Gl~581 having 
metallicities of respectively [Fe/H]=$-$0.03, $+$0.02 and $-$0.25~dex
\citep{Bonfils2005}. This contrasts with the median metallicity for 
solar-type stars surrounded by planets, [Fe/H]=$+$0.2 
\citep{2005A&A...437.1127S}, though the significance of the difference
is obviously still modest.

As discussed in the introduction, various groups monitor over 200 M~dwarfs 
with 3-15~m~s$^{-1}$ precision,
sufficient to easily detect the $>$40~m~s$^{-1}$ reflex motion of a 
0.3M$_{\odot}$ star orbited by a Jupiter-mass planet out to 2~AU.
That these efforts have to date found only 5 planets, of which only
Gl~876b and c have approximately jovian masses, demonstrates that there
are much fewer ${\approx}M_{Jup}$ planets around M~dwarfs than the
$\approx$5\% \citep{2000prpl.conf.1285M,naef05} found around solar-type stars.
The 5 planets include no hot-Jupiter, but with only $\sim$1\% 
solar-type star orbited by such a planet the significance of that fact is 
still modest. 3 of the 5 on the other hand are hot-Neptunes (Gl~436b, 
M~sin~i=1.2~M$_{Nep}$; Gl~876d, 0.44~M$_{Nep}$; Gl~581b, 0.99~M$_{Nep}$), 
as many as currently known around all solar-type stars. This matches the 
theoretical model of \cite{2005ApJ...626.1045I}: 
the mass-distribution of close-in planets has two peaks centered
at about the masses of Jupiter and Neptune, with the former preferentially
populated around G-dwarfs and the latter around M dwarfs, reflecting
how much matter remains available in the disk for accretion during
the inward migration of the planet. Other theoreticians however take the 
view that many hot-Neptunes are actually evaporated hot-Jupiters 
\citep{2005A&A...436L..47B}. Better statistics on M-dwarf planets will 
help determine which of these mechanisms dominate. 

A final striking characteristic of the current M-dwarf planets is 
that none has a period longer than the 2~months of Gl~876b. 
By contrast, 66\% of the 164 planets known around solar-type stars have 
orbital periods above 100~days and their distribution is even observed to increase with period \citep{udry03}. 
With 5~planets known around M~dwarfs, the
probability that the long-period deficit amongst M-dwarf occurs by chance
is thus less than (0.34)$^5$=5.$10^{-3}$. The sensitivity of 
Doppler searches does degrade for longer periods however, and planets of 
M dwarfs have often been found close to the sensitivity floor of their 
respective discovery surveys, making detection biases potentially important.
A full account is not currently possible from published information (our
own high precision M-dwarf survey is still too recent to be very useful in 
this respect), but for periods shorter than the observing interval the 
sensitivity degrades only slowly ($\propto$P$-1/3$). Some of the 
M-dwarf Doppler surveys have been observing for long enough 
\citep{2004ASPC..321..101B,2004ApJ...617..580B} 
that detection biases make an unlikely full explanation for the
lack of long period planets. That deficit therefore has to be at least
in part intrinsic, perhaps reflecting smaller protoplanetary disks around 
the lower mass M dwarfs.

\begin{acknowledgements}
We thank our technical and scientific collaborators of the HARPS Consortium, 
ESO Head Quarter, and ESO La Silla, who have contributed with passion and 
competence to the success of the HARPS project. We are also grateful to 
Damien S\'egransan who contributed additional ELODIE observations, and to 
Jean-Christophe Leyder and collaborators for using 
some of their own observing time to obtain critical confirmation observations.

\end{acknowledgements}

\begin{table}
\caption{Radial-velocity measurements and error bars for Gl\,581. All 
values are relative to the solar system barycenter. Only available
electronically}
\label{TableRV}
\centering
\begin{tabular}{c c c}
\hline\hline
\bf JD-2400000 & \bf RV & \bf Uncertainty \\
 & \bf [km~s$^{-1}$] & \bf [km~s$^{-1}$] \\
\hline
53152.71289&  -9.2235 & 0.0012\\
53158.66336&  -9.2319 & 0.0014\\
53511.77355&  -9.2202 & 0.0014\\
53520.74475&  -9.1999 & 0.0016\\
53574.52223&  -9.2024 & 0.0013\\
53575.48075&  -9.2069 & 0.0011\\
53576.53646&  -9.2202 & 0.0011\\
53577.59250&  -9.2221 & 0.0014\\
53578.51061&  -9.2108 & 0.0010\\
53578.62960&  -9.2092 & 0.0013\\
53579.46256&  -9.1991 & 0.0011\\
53579.62115&  -9.1970 & 0.0012\\
53585.46167&  -9.2034 & 0.0013\\
53586.46516&  -9.2141 & 0.0010\\
53587.46481&  -9.2269 & 0.0020\\
53588.53827&  -9.2179 & 0.0036\\
53589.46202&  -9.2051 & 0.0010\\
53590.46379&  -9.1990 & 0.0010\\
53591.46638&  -9.2034 & 0.0010\\
53592.46481&  -9.2175 & 0.0010\\
\hline
\end{tabular}
\end{table}

\end{document}